\documentstyle[eqsecnum,multicol,epsfig,aps,prl,array]{revtex}
\newcommand{\bleq}{\ifpreprintsty
                   \else
                   \end{multicols}\widetext \vspace*{-3.5ex}{\tiny
                   
                \noindent\begin{tabular}[t]{c|}
                   \parbox{0.493\hsize}{~} \\ \hline \end{tabular}}
                                      \fi}
\newcommand{\eleq}{\ifpreprintsty
                   \else
                   {\tiny\hspace*{\fill}\begin{tabular}[t]{|c}\hline
                    \parbox{0.49\hsize}{~} \\
                    \end{tabular}}\vspace*{-2.5ex}\begin{multicols}{2}
                    \narrowtext
                    \fi}
\newcommand{\bcols}{\ifpreprintsty\else\begin{multicols}{2} 
        \narrowtext\fi}
\newcommand{\ecols}{\ifpreprintsty\else\end{multicols}\fi}

\widetext
\begin{document}
\title{The critical behaviour of the 2D Ising model in Transverse
  Field; a Density Matrix Renormalization calculation.} \author{M.  S.
  L. du Croo de Jongh and J. M. J. van Leeuwen}
\address{Instituut-Lorentz, Leiden University, \\ P. O. Box 9506, 2300
  RA Leiden, The Netherlands}

\date{\today} \maketitle
\begin{abstract}
  We have adjusted the Density Matrix Renormalization method to handle
  two dimensional systems of limited width. The key ingredient for
  this extension is the incorporation of symmetries in the method.
  The advantage of our approach is that we can force certain symmetry
  properties to the resulting ground state wave function.  Combining
  the results obtained for system sizes up-to $30 \times 6$ and finite
  size scaling, we derive the phase transition point and the critical
  exponent for the gap in the Ising model in a Transverse Field on a
  two dimensional square lattice.

\end{abstract}
 \bcols
\section{Introduction}

The calculation of ground state properties of a quantum system with
many degrees of freedom has been explored by several means. Exact
diagonalisation is usually limited to fairly small sized systems.
Monte Carlo methods are hard to extend to zero temperatures and/or are
seriously hampered by sign problems in the wave function (in
particular for fermionic degrees of freedom). Recently White
\cite{white92} has introduced a new algorithm, which bears some
analogy with the renormalization technique in the sense that wave
functions of larger systems are constructed hierarchically from
smaller components.  It has received the name Density Matrix
Renormalization Group (DMRG) although the group character is nowhere
present and even the link with renormalization as induced by spatial
rescaling is rather weak. The DMRG-method has achieved remarkable
accuracy for a number of systems notably those of a 1-dimensional
($d=1$) character. Although there is a physically acceptable rational
for the algorithm, its limitations are not well understood. In
particular the restriction of the success to $d=1$ systems is of an
empirical nature while theoretically the renormalization idea would
equally well work in higher dimensions.  As noted earlier the method
performs poorest \cite{drzewinski94,ostlund95} near a quantum phase
transition. According to \"Ostlund and Rommer \cite{ostlund95} this
has to do with the hierarchical nature of the ground state wave
function which is at odds with the algebraic correlations in a
critical system.

In this paper we study the Ising model in a Transverse Field (ITF) as
a model system for a quantum phase transition on a 2 dimensional
square lattice. It has the advantage to be exactly soluble in $d=1$
dimensions which helps checking accuracies for the method we use.
Straightforward application of the DMRG-methods yields highly accurate
results in this case. The real challenge is $d=2$ dimensions where the
phase transition has the same complexity as that of a $d=3$
dimensional classical Ising model. Here straightforward, brute force
application of the DMRG-technique does not yield convincing results
and sophistication is called for. Still some results on $d=2$
dimensional spin system have been achieved by White \cite{white96}.

\begin{figure}
\begin{center}
  \epsfig{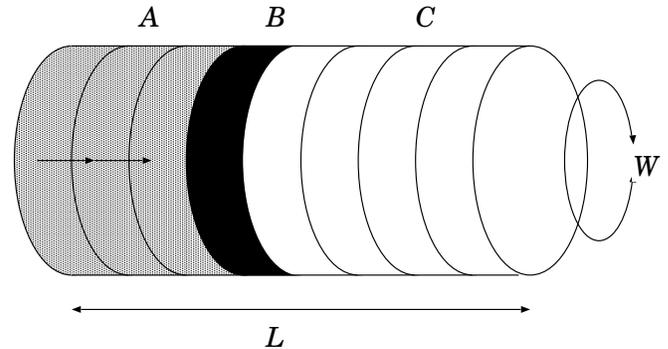}
\caption{The systems we consider are of dimensions $L \times W$ where
  $L=2W,3W,5W,20$ and $W=2, \dots ,6$. The system contains three
  parts: A left-hand part $A$ (shaded), an intermediate band $B$
  (black) and a right-hand part $C$ (white). Actually the figure
  obscures that for $L=2W,3W,20$ the systems are periodical in both
  directions. At every DMRG step part $A$ and $B$ are contracted.}
\label{fig:system}
\end{center}
\end{figure}

The new ingredient in this paper is that we grow the system with a
whole band per step in stead of just one site (see figure
\ref{fig:system}). This approach allows us to implement the
translational symmetry in the $W$-direction in a similar fashion as
Xiang \cite{xiang96} has done for the Hubbard model.

The lay-out of this paper is as follows: First we introduce the ITF.
The critical behaviour of this model is discussed next. After that, we
shift our attention to the DMRG. We make a link with perturbation
theory and show the limitations of the method as a consequence of the
environment, part $C$. Afterwards our implementation is described.
Finally, we present the results of the calculations: On one hand, the
accuracies achieved and on the other hand, the critical properties of
the ITF.

\section{The ITF}

Since the beginning of the 1960s, the ITF has been studied. In first
instance, the purpose was to model specific materials like $KH_2PO_4$
crystals. With the introduction of the renormalization group in the
1970s, an other interest in this model has arisen. As there exist a
simple relation to a classical system, the ITF is used as a vehicle to
extend the knowledge of critical phenomena from classical spin systems
to quantum systems. Chakrabarti, Dutta and Sen \cite{chakrabarti} have
recently summarised the properties of the ITF. For further details we
refer to them. We will only mention those properties here, that are of
explicit use to our calculations.

\subsection{The Model}

Consider a two dimensional square lattice with length $L$ and width
$W$. The lattice is periodic in both directions and each lattice site
contains a spin-$\frac{1}{2}$. The Hamiltonian is given by
\begin{equation}
  {\cal H}_{ITF} = \sum_{i=1}^{L} \sum_{j=1}^{W} \left ( -J {\cal
    S}_{i,j}^{x} ({\cal S}_{i+1,j}^{x} + {\cal S}_{i,j+1}^{x} ) + H
  {\cal S}_{i,j}^z \right )
\end{equation}
where the ${\cal S}^{\alpha}_{i,j}$ are the usual Pauli spin matrices
satisfying
\begin{equation}
  [{\cal S}_{i,j}^{\alpha},{\cal S}_{i',j'}^{\beta}] = 2i
  \delta_{i,i'} \delta_{j,j'} \epsilon_{\alpha \beta \gamma} {\cal
    S}_{i,j}^{\gamma} ~~,~~ \alpha,\beta,\gamma=x,y,z
\end{equation}

As only the ratio $H/J$ is important, we fix $J$ at $J=1$ and take
$H\ge 0$ ( $H \le 0$ being equivalent).

This model is translation and reflection symmetric in both directions.
Moreover, the symmetry operation ${\cal S}_{i,j}^x \rightarrow -{\cal
  S}_{i,j}^x$, ${\cal S}_{i,j}^y \rightarrow -{\cal S}_{i,j}^y$ and
${\cal S}_{i,j}^z \rightarrow {\cal S}_{i,j}^z$ leaves the model
invariant. The operator associated with this is ${\cal S}= \exp(i
\pi/2 (\sum_{i,j} {\cal S}_{i,j}^z + LW))$.  It samples the total
number of spins pointing upwards and returns whether it is odd ${\cal
  S}=-1$, or even ${\cal S}=+1$. We call this ${\cal S}$ the
spin-reversal operator. The ground state is an eigenfunction of all
these symmetries operations.

If $H \rightarrow 0$, we end up with a simple 2D Ising model. The
ground state is degenerate; all spins point either up or down in the
${\cal S}^x$-direction. The associated phase is the classical ordered
phase. By taking a rotation in the lowest energy space, we can obtain
states that are even in spin-reversal (${\cal S}=+1$) or odd (${\cal
  S}=-1$). In the other extreme, $1/H \rightarrow 0$, free spins in an
external field remain.  The ground state is unique and has all spins
pointing down in the ${\cal S}^z$-direction. This is the reference
state for the quantum disordered phase and has value ${\cal S}=+1$.
The lowest excitation differs from the ground state by the reversal of
one spin. So it belongs to the class ${\cal S}=-1$. We will
extensively study the energy gap $\Delta$ between the lowest
excitation (in ${\cal S}=-1$) and the ground state (in ${\cal S}=+1$);
$\Delta = E_{ex}-E_{gr}$.

There is a phase transition between the classical ordered and the
quantum disordered state. A clear signature of this phase transition
is the disappearance of the gap $\Delta$, which occurs for a critical
value $H=H_c$.

\subsection{Critical Behaviour}

As mentioned before, the ITF is closely related to a classical Ising
model. It can be mapped onto an anisotropic Ising model in one
dimension higher. In the current situation the resulting classical
model is of size $L \times W \times \infty$. It contains a weak
coupling $K_{\perp}$ in the $L \times W$-plane and a strong Ising
coupling $K_{||}$ in the remaining direction ($\exp
(-K_{||})=\varepsilon H$, $K_{\perp}=\varepsilon J$ with $\varepsilon
\ll 1$). Chakrabarti et al.  \cite{chakrabarti} give an overview of
the procedure. For our purposes the most important consequences are:
\begin{itemize}
\item The correlation length $\xi$ in the strong coupling direction
  corresponds to the inverse of our gap $\Delta$ ($\xi \sim
  \Delta^{-1}$).
\item The reduced temperature $t=(T-T_c)/T_c$ corresponds to our
  reduced field $h=(H-H_c)/H_c$. ($t \sim h$).
\end{itemize}
The well-known relation $\xi \sim t^{-z\nu}$ transforms into $\Delta
\sim h^{z\nu}$. In the 3D anisotropic Ising model the dynamical
exponent $z=1$. We use this in the further discussion. The standard
finite size scaling methods as described in \cite{cardy} can be
applied here. The classical scaling relation $\xi(t,W^{-1})=b
\xi(b^{1/ \nu} t,bW^{-1})$, for fixed aspect ratio, $L/ W =$ constant,
becomes
\begin{equation}
  \Delta(h,W^{-1})=\frac{1}{b} \Delta(b^{1/\nu} h,b W^{-1}). \label{eq:scal1}
\end{equation}
We may set $b=W$ and obtain the scaling expression
\begin{equation}
  W\Delta(h,W^{-1})=\Delta(W^{1/\nu} h,1), \label{eq:scaling}
\end{equation}
showing that $W\Delta$ only depends on the combination $W^{1/\nu}h$.
So for $h=0$ all lines $W\Delta$ cross at the same value;
\begin{equation}
  W\Delta(0,W^{-1})=\Delta(0,1) \label{eq:Hc}
\end{equation}
This gives us $H_c$. If we differentiate (\ref{eq:scaling}) with
respect to $h$ and set $h=0$ afterwards, we obtain
\begin{equation}
  \left ( 1 -\frac{1}{\nu} \right ) \log \left ( \frac{W}{W+1} \right
    ) = - \log \left (
    \frac{\Delta_h(0,W^{-1})}{\Delta_h(0,(W+1)^{-1})} \right ).
    \label{eq:nu}
\end{equation}
From this we can extract $\nu$.

\section{the DMRG Method}

The DMRG method was first formulated by White \cite{white92}. As it is
not a renormalization group method in the traditional sense, it could
perhaps better be named an iterative basis truncation method. Gehring,
Bursill and Xiang \cite{gehring96} provide an excellent introduction
in the application of the DMRG to 1D spin systems. Here we will not be
so extensive. Two important features of the method are discussed and
our approach is outlined.

\subsection{Limitations by the Environment}

The essence of the DMRG can be described as follows: Consider a system
consisting of 2 parts, $A$ and $B$. Moreover, suppose we have an
approximate ground state wave function of the combined system $|
\phi_0 \rangle = \sum_{ij} \phi_{ij} |i \rangle |j \rangle$. The bases
$\{ |i \rangle \}$ and $\{ | j \rangle \}$ do not have to be complete.
We want to reduce the number of basis states in part $A$, preserving
the ground state wave function $|\tilde{\phi}_0 \rangle$ as well as
possible. $|\phi_0 \rangle$ can be expanded as
\begin{equation}
  |\tilde{\phi}_0 \rangle = \sum_{\alpha j} \tilde{\phi}_{\alpha j}
  |\alpha\rangle |j \rangle.
\end{equation}
where $\{ |\alpha \rangle \}$ spans only a subspace of $\{ |i \rangle
\}$.  Preserving the ground state means that
\begin{equation}
  \left | | \phi_0 \rangle - | \tilde{\phi}_0 \rangle \right|^2 \hbox{
      is minimal.}
\end{equation}

The solution to this problem can be obtained by means of simple
algebra \cite{white92}; Construct the density matrix $\rho_{ii'}
=\sum_j \phi_{ij} \phi_{i'j}$ and select the eigenvectors
$\vec{\nu}^{\alpha}$ with the largest eigenvalues $\lambda^{\alpha}$;
$\vec{\vec{\rho}} \cdot \vec{\nu}^{\alpha} = \lambda^{\alpha}
\vec{\nu}^{\alpha}$. The new basis is now given by: $| \alpha \rangle
= \sum_i \nu^{\alpha}_i |i \rangle$.

A truncation error
\begin{equation}
  p = 1- \sum_{\alpha}\lambda^{\alpha} = \left | |\phi_0 \rangle -|
  \tilde{\phi}_0 \rangle \right |^2
\end{equation}
is introduced to give an indication of the effectiveness of the
procedure. On basis of experience this truncation error is said to be
a measure of the error in the calculated energy with respect to the
exact result.

There is a peculiarity that was only briefly mentioned by White
\cite{white92}.  Suppose we want to use this selection scheme to
obtain as many states in part $A$ as we already have in part $B$,
presuming that there were more states in $A$ initially. Define $|
\beta_j \rangle = \sum_i \phi_{ij} | i \rangle$. The ground state can
be transformed into this set;
\begin{equation}
  | \phi_0 \rangle = \sum_{ij} \phi_{ij} |i\rangle |j\rangle = \sum_j
  | \beta_j \rangle |j \rangle
\end{equation}
Thus by orthonormalising the set $\{ | \beta_j \rangle \}$ we obtain a
basis set for $A$ in which the wave function can exactly be
reproduced.  A reformulation of this is: Consider a $|\alpha \rangle $
such that $ \langle \beta_j| \alpha \rangle =0$ for all $| \beta_j
\rangle$.  We know that $| \alpha \rangle = \sum_{i} \nu_i^{\alpha} |
i\rangle $, thus
\begin{equation}
  \langle \beta_j| \alpha \rangle=\sum_{i'} \phi_{i'j}
  \nu^{\alpha}_{i'} = 0
\end{equation}
and
\begin{equation}
  \sum_{i'}\rho_{ii'} \nu_{i'}^{\alpha} = \sum_{i'j} \phi_{ij}
  \phi_{i'j} \nu^{\alpha}_{i'} = 0.
\end{equation}
$\vec{\nu}^{\alpha}$ is a zero eigenvector of $\vec{\vec{\rho}}$.
Keeping the subspace spanned by the $|\beta_j \rangle$ would make the
truncation error $p$ equal to zero. This lack of choice only becomes
worse in case symmetries are implemented; not just the total number of
non-zero eigenstate is fixed, but even within a specific symmetry
class the number of non-zero eigenstate is dictated by the states in
the environment. Later on, we will make this explicit for the systems
we consider.

\subsection{The Connection with Perturbation Theory}

Liang and Pang \cite{liang94} mention that for a given accuracy, the
number of states needed in a single part of the system grows
exponentially with the width of the system. At the phase transition,
we confirm this observation (figure \ref{fig:error}).  Moreover, both
in the weak and strong field limits ($H \ll 1$ and $1/H \ll 1$), we
find a fast convergence, which can be explained by perturbation
theory.
 
Consider the quantum disordered phase. Split the Hamiltonian into a
unperturbed part ${\cal H}_0=H\sum_{i,j} {\cal S}_{i,j}^z$ and a
perturbation ${\cal V}=- \sum_{i,j} {\cal S}_{i,j}^x ({\cal
  S}_{i+1,j}^x+ {\cal S}_{i,j+1}^x)$. We split the periodical,
rectangular system of size $L\times W$ again in two parts; $A$ and $B$
of sizes $l\times W$ and $(L-l)\times W$ where $l$ is an arbitrary
length smaller than $L$.  They both contain $2W$ spins that border the
other part. The unperturbed ground state $|0\rangle $ has all spins
pointing down in the ${\cal S}^z$-direction. It is the direct product
of two equivalent states restricted to $A$ and $B$; $|0 \rangle = |0
\rangle_A |0\rangle_B$. We know that ${\cal H}_0|0 \rangle =
-HLW|0\rangle=E_{0}|0 \rangle$.  Perturbation theory yields
\begin{equation}
  |\phi_0 \rangle = |0\rangle + \frac{1}{E_{0}-{\cal H}_0} {\cal V} |0
  \rangle + {\cal O}\left (\frac {1}{H^2} \right )
\end{equation}
The perturbation flips a pair of neighbouring spins. This pair can be
in a single part or it can cross the border between both parts. In the
latter case the spins are adjacent across the boundary between part
$A$ and $B$. Define $\{ |a \rangle_A \}$ to be the set of states with
the flipped pair in part $A$. Idem for $\{ | b \rangle_B \}$. Moreover
let $\{ |n \rangle_A \}$ be the set with one spin flipped on the $n$th
boundary site with $B$ and define in an equivalent manner $\{ |n
\rangle_B \}$. The perturbation expansion can now be rewritten \bleq
\begin{eqnarray}
  | \phi_0 \rangle &=& | 0 \rangle_A |0\rangle_B + \frac{1}{2H} \left
  ( \sum_a |a \rangle_A |0 \rangle_B + \sum_b |0 \rangle_A |b
  \rangle_B +\sum_n |n\rangle_A |n \rangle_B \right)+{\cal O} \left(
  \frac{1}{H^2} \right ) \nonumber \\ &=& \left ( |0 \rangle_A +
  \frac{1}{2H} \sum_a |a \rangle_A \right) \left ( |0 \rangle_B +
  \frac{1}{2H} \sum_b |b \rangle_B \right ) + \frac{1}{2H} \sum_n
  |n\rangle_A |n \rangle_B + {\cal O} \left(\frac{1}{H^2} \right )
  \label{eq:pert}
\end{eqnarray}
\eleq As $H \gg 1$, it is necessary to reproduce {\it all} these terms
for an accuracy which is equivalent to the first order perturbation
theory. The minimal number of states needed in part $A$ is therefore
$1$ for the first term in (\ref{eq:pert}) plus $2W$ for all the
boundary terms. We have confirmed this prediction explicitly in both
the small and large $H$ limit ( $H=1/50,50$).

The same line of reasoning also holds for the second and higher order
perturbation terms. We expect for an error comparable to the $n$th
order perturbation theory that $m \sim W^n$, $\delta E \sim (1/H)^n$.
This is always an upper bound for number of states $m$ needed, $m<W^n$
for a given accuracy $\delta E \sim (1/H)^n $. Only when the different
orders in perturbation theory become distinguishable in size - the
limit of large $H$- the equivalence holds.

\subsection{Exploiting the Symmetries}

We consider systems of sizes $L\times W$. The length $L$ is either a
multiple of $W$, $L=2W,3W,5W$, or it is fixed, $L=20$. The width $W$
is varied from $W=2$ to $W=6$. The maximal system we study thus
contains $6*30=180$ spins. For $L=2W,3W$ and $20$ a torus is constructed by
imposing periodical boundary conditions in both directions. For $L=5W$
the system follows the figure \ref{fig:system} more genuinely; it is
periodical in the width-direction and open in the length-direction.
The system is split in a left-hand and a right-hand part, both
containing $m$ states. A intermediate band, containing the complete
basis of $2^W$ states, separates them.  This is depicted in
figure \ref{fig:system}.

The Hamiltonian of such a system contains many symmetries that we can
incorporate in our calculation. The general form of the included
symmetry operators is that they are the direct product of three
components.  Each component acts on one part of the system only. For
example, consider the translation operator ${\cal T}$ in the
width-direction.  This operator is the direct product of three
translations in the individual parts; ${\cal T}={\cal T}_A {\cal T}_B
{\cal T}_C$. The same holds for the reflection ${\cal R}$ in the same
direction,${\cal R}={\cal R}_A {\cal R}_B {\cal R}_C$, and the
spin-reversal operator ${\cal S}=\exp(i\pi /2 (\sum_{i,j} {\cal
  S}^z_{i,j}+LW))={\cal S}_A {\cal S}_B {\cal S}_C$.

The ground state $| \phi_0 \rangle$ of the system is translational,
reflection and spin-reversal invariant; ${\cal T}| \phi_0
\rangle={\cal R}| \phi_0 \rangle={\cal S}| \phi_0 \rangle=| \phi_0
\rangle$.  For systems of infinite size in the classical ordered
region ($L,W \rightarrow \infty$ and $ H \ll 1$), it will become
degenerate with a state that is spin-reversal anti-symmetric.  In
order to take advantage of the symmetries, the bases of part $A$, $B$
and $C$ are chosen to be eigenvectors of the symmetry operators ${\cal
  T}$ and ${\cal S}$.  ${\cal R}$ is used later on. So if $\{ |a
\rangle \}$, $\{ |b \rangle \}$, $\{ |c \rangle \}$ are the bases of
the individual parts ( not to be mixed-up with the notation used in
(\ref{eq:pert})) then
\begin{equation}
  {\cal T}_A |a \rangle = e^{ik_a}|a \rangle ~~,~~ {\cal S}_A |a
  \rangle=s_a |a \rangle. \label{eq:symm}
\end{equation}
Similar relations hold for the other two sets. Thus
\begin{equation}
  |\phi_0 \rangle = \sum_{abc} \phi_{abc} |a \rangle |b \rangle |c
  \rangle
\end{equation}
and application of the symmetry operations together with
(\ref{eq:symm}) yields:
\begin{eqnarray}
  k_a+k_b+k_c&=&0 ~\hbox{mod}~ 2\pi, \\ s_a s_b s_c&=&1.
\end{eqnarray}
It is also possible to set up the program to find the lowest state in
other symmetry classes by forcing other values than $0$ and $1$ in the
equations above.

The Hamiltonian can be written as the sum of Hamiltonians within the
separate parts: ${\cal H}_A,{\cal H}_B$ and ${\cal H}_C$ combined with
interactions between parts: ${\cal H}_{AB},{\cal H}_{BC}$ and ${\cal
  H}_{CA}$; ${\cal H}={\cal H}_A+{\cal H}_B+{\cal H}_C+{\cal
  H}_{AB}+{\cal H}_{BC}+{\cal H}_{CA}$. To show how to implement the
symmetries, we will discuss one element of both types.

First ${\cal H}_A$: It is translational and spin-reversal invariant,
thus
\begin{eqnarray}
  \langle a'| {\cal H}_A |a \rangle &=& \langle a'| {\cal
    T}_A^{-1}{\cal H}_A {\cal T}_A|a \rangle = e^{i(k_a-k_{a'})}
  \langle a'|{\cal H}_A|a \rangle \nonumber \\ &=& \langle a'|{\cal
    S}_A^{-1}{\cal H}_A{\cal S}_A|a \rangle=s_{a'}s_a\langle a'|{\cal
    H}_A|a \rangle \nonumber \\ &=& \langle a'| {\cal H}_A |a \rangle
  \delta_{s_{a'},s_a} \delta_{k_{a'},k_a}.
\end{eqnarray}
It only contains elements within symmetry classes, as one would
expect.

Second ${\cal H}_{AB}$: Once again, it is translational and
spin-reversal invariant.  Moreover it can be written as
\begin{eqnarray}
  {\cal H}_{AB} &=&-\sum_{n=1}^{W} {\cal S}^x_{l,n} {\cal S}^x_{l+1,n}
  \nonumber \\ &=&-\sum_{n=1}^{W} ({\cal T}_A{\cal T}_B)^{-n+1} {\cal
    S}^x_{l,1} {\cal S}^x_{l+1,1} ({\cal T}_A{\cal T}_B)^{n-1}
  \label{eq:Hab}
\end{eqnarray}
where $l$ is the length of part $A$. ${\cal S}_{i,j}^x$ flips a spin,
so $ {\cal S}_{i,j}^x{\cal S} + {\cal S}{\cal S}_{i,j}^x=0$. Inserting
this and (\ref{eq:symm}) in (\ref{eq:Hab}) gives
\begin{eqnarray}
  \langle a'| \langle b'|{\cal H}_{AB}|a\rangle |b \rangle &=& -W
  \langle a'| {\cal S}_{l,1}^x |a \rangle \langle b' | {\cal
    S}_{l+1,1}^x |b \rangle \cdot \nonumber
    \\ & & \delta_{k_{a'}+k_{b'},k_a+k_b} \cdot \nonumber \\
    & & \delta_{s_{a'},-s_a} \delta_{s_{b'},-s_b}.
\end{eqnarray}
This substantially reduces the computational effort. Finally: the
reflection operator ${\cal R}$ is used to make matrix elements like $
\langle a'| {\cal S}_{l,1}^x |a \rangle$ real. Naturally we could have
used this last symmetry ${\cal R}$ more, but it only reduces the
effort by a factor of 4 while making the program far more complex.

\subsection{The Implementation}

Now we focus on the procedure itself. It is tempting to use the 1D
DMRG method directly: a site is replaced by a band. The ground state
$| \phi_0 \rangle$ of the entire system ABC is calculated and the
optimal basis for block AB is selected through the density matrix.
However, one runs in severe difficulties as a consequence of the first
remark on the DMRG. It is instructive to reveal the reason: Using the
notation above, we define $|\beta_c \rangle= \sum_{ab} \phi_{abc}
|a\rangle |b \rangle$. We know that ${\cal T} |\phi_0 \rangle ={\cal
  S}|\phi_0 \rangle=|\phi_0\rangle$, thus
\begin{eqnarray}
  {\cal T}_A {\cal T}_B | \beta_c \rangle &=& e^{-ik_c} |\beta_c
  \rangle \nonumber \\ {\cal S}_A {\cal S}_B | \beta_c \rangle &=& s_c
  | \beta_c \rangle
\end{eqnarray}
The distribution over the symmetry classes in part $C$ forces the
selected states in block $AB$ to be in ``conjugate''-classes. To
overcome this problem, we need to increase the number of states in
part $C$. In that case we can really make a selection and shift into
important symmetry classes.

In the 1D procedure the solution is to add one extra site to the
environment. The number of states in the environment is then doubled.
In our set-up this would correspond to adding an extra band between B
and C. This is computational far too expensive.  We now introduce
variants on White's infinite-size and finite-size algorithms
\cite{white92} that increase the number of states in the part C.

\begin{figure}
\begin{center}
  \epsfig{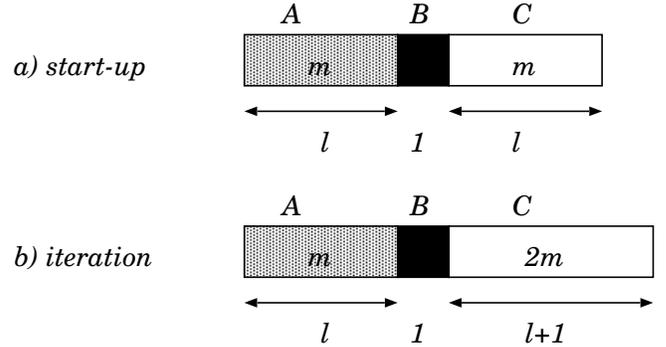}
\caption{A inductive step in the infinite-size procedure consists of
  a start-up to obtain an initial approximation for states in a system
  of length $l+1$ and iterative calculations to make the basis
  converge. The numbers in the rectangle are the number of states in
  the parts. The intermediate band B always contains the complete
  basis of $2^W$ states.}
\label{fig:infinite}
\end{center}
\end{figure}

First we consider our infinite size approach. We only have to describe
one step in the process as it is an inductive method. We have a basis
of $m$ states for a system of length $l$.
\begin{itemize}
\item We construct the combined system as depicted in figure
  \ref{fig:infinite}-a by taking this basis in part $A$ and $C$
  together with the complete basis in the intermediate band $B$.
  ($L=2l+1$)
\item We calculate the ground state $|\phi_0 \rangle$ and obtain $m$
  basis states for a system of length $l+1$ by orthonormalising $\{
  |\beta_c \rangle \}$.
\item Suppose that block $AB$ has $f$ symmetry classes. To every
  symmetry class we add $m/f$ basis states constructed {\it randomly}
  from the $m2^W$ states in $A$ and $B$. We end up with $m+f\cdot
  m/f=2m$ basis states for a system of length $l+1$.
\item In part $A$ we now take the $m$ basis states for a system of
  length $l$ and in part $C$ we take the newly constructed $2m$ states
  for length $l+1$. ($L=2l+2$) This yields the configuration in figure
  \ref{fig:infinite}-b.
\item We calculate the ground state $|\phi_0 \rangle$ and obtain $2m$
  basis states for length $l+1$ by orthonormalising $\{ |\beta_c
  \rangle \}$. We replace the basis of part $C$ by this basis and
  repeat this step a couple of times ($\sim 3$).
\item We {\it select} from the $2m$ basis states for length $l+1$ $m$
  states on basis of the density matrix.
\end{itemize}
Now we have returned to the original situation with the exception that
$l$ has increased by one. The new ingredient is thus to add $m$ random
states to the basis and iterate until the result has converged.

\begin{figure}
\begin{center}
  \epsfig{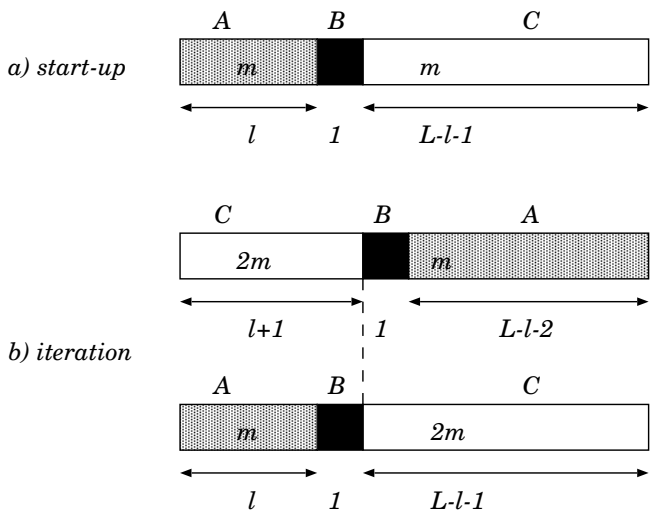}
\caption{A inductive step in the finite-size procedure also consists
  of a start-up to obtain an initial approximation for states in a
  system of length $l+1$. Afterwards we move back and forth between
  lengths $l$ and $l+1$ to make this converge. }
\label{fig:finite}
\end{center}
\end{figure}

In the same line our finite size approach lies. Suppose we have basis
sets of $m$ states for lengths $l$,$L-l-1$ and $L-l-2$, where $L$ is
now fixed and independent of $l$.
\begin{itemize}
\item We take the basis for $l$ in part $A$, the basis for $L-l-1$ in
  part $C$ and the complete basis of the band in part $B$. See figure
  \ref{fig:finite}-a.
\item We calculate the ground state $| \phi_0 \rangle $ and obtain a
  basis for length $l+1$ by orthonormalising $\{ |\beta_c \rangle \}$.
\item In the same way as in the infinite-size algorithm we add $m$
  randomly chosen states to this basis.
\item In part $C$ we take the $2m$ basis states for length $l+1$ and
  in part $A$ the $m$ states for length $L-l-2$. This is depicted in
  the first of the two pictures in figure \ref{fig:finite}-b.
\item We calculate the ground state $|\phi_0 \rangle$ and obtain $2m$
  basis states for $L-l-1$.
\item In part $C$ we take the $2m$ basis states for length $L-l-1$ and
  in part $A$ the $m$ states for length $l$; see the second picture in
  figure \ref{fig:finite}-b.
\item We calculate the ground state $|\phi_0 \rangle$ and obtain $2m$
  basis states for $l+1$. These last four steps are repeated a couple
  of times ($\sim 3$)
\item We {\it select} from the $2m$ basis states for length $l+1$ $m$
  states on basis of the density matrix.
\end{itemize}
Once again we have returned to our starting position while increasing
the length $l$ by one. By sweeping through the system we can therefore
systematically improve the basis. This method convergences at a
similar speed as the 1D approach; After 3 sweeps through the system
the final result is achieved.

The computational effort scales as $m^3 L \frac{2^{W}}{W}$. In general
$m \gtrsim 2^W$.  This clarifies the bound on the width. The
alternative is to follow Liang, Pang \cite{liang94} and White
\cite{white96} by adding one site per step.  We can then still use
${\cal S}$. The calculation would scale as $ m^3 L \frac{W^2}{2^2}$.
Our approach includes the symmetry requirements of the ground state
and up to $W \sim 8$ it is similar in speed as theirs. Applying our
method to models where the number of particles or total spin is
conserved instead of ${\cal S}$, the calculations can been
substantially reduced and systems of widths $W > 6$ are pulled within
reach.

The largest calculation, $W=6$, $L=18$, $m=200$ took 48h of computer
time per $H$-value (at 462 SPECfp92). To determine the gap $\Delta$
two such points are needed (${\cal S}=+1$ and ${\cal S}=-1$).

\section{Results}

We have performed two kinds of calculations: First, we made a check on
the accuracy of the method. Second, we have calculated the gap
$\Delta$ for various widths $W$, aspect ratio's $L=xW$ and fields $H$
in order to find through finite size scaling the phase transition
point $H_c$ and the critical exponent $\nu$.

\subsection{The Accuracy of the Energies.}

\begin{figure}
\begin{center}
  \epsfig{figure=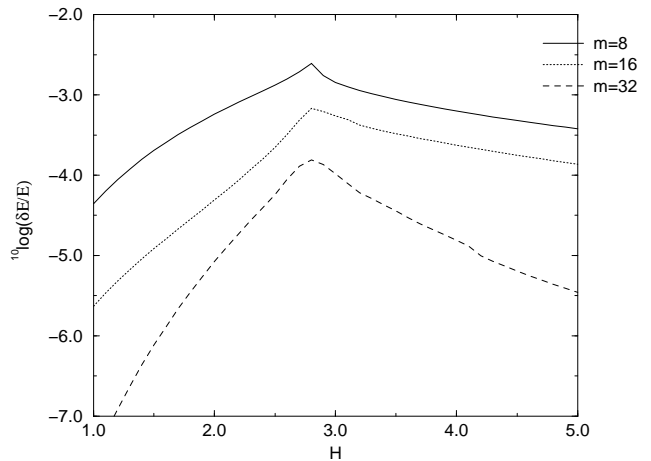,width=\linewidth}
\caption{The accuracy of the DMRG method for different of states $m$ 
  (numbers in graph) as function of the field $H$. The system is
  periodical in both directions with dimensions $W=4$ and $L=20$.  The
  reference value is taken from a DMRG calculation with $m=64$.}
\label{fig:errorHz}
\end{center}
\end{figure}

The strict method to determine the error in the energy $\delta E_m$
for given number of states $m$ is to compare the results $E_m$ with
the exact value $E_{gr}$; $\delta E_m=E_m-E_{gr}$. This would limit us
to small systems of sizes comparable to $6 \times 6$. In the
literature \cite{liang94} it is noted that the error $\delta E$
decreases exponentially with the number of states $m$ included. We
confirm that statement explicitly for these small systems. Moreover we
use this feature to test the accuracy for far larger systems. The
energy $E_m$ is compared with the result for a larger number of
states. For instance $m<128$; $\delta E_m \approx E_m-E_{128}$. The
error $\delta E_m$ is largest near the phase transition as can be seen
in figure \ref{fig:errorHz}.

\begin{figure}
\begin{center}
  \epsfig{figure=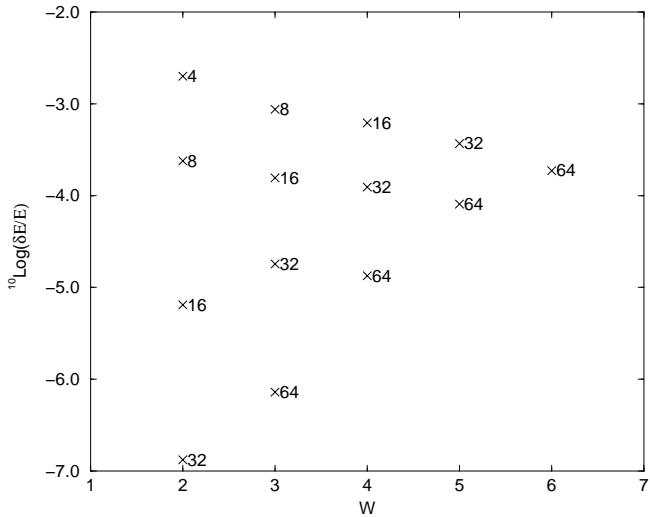,width=\linewidth}
\caption{The accuracy of the DMRG method for given number of states $m$ 
  (numbers in graph) as function of the width $W$. $H=3$ and $L=20$.
  The system is periodical in both directions.  The reference value is
  taken from a DMRG calculation with $m=128$.}
\label{fig:error}
\end{center}
\end{figure}

As the phase transition occurs near $H=3$, we take $H=3$ as an example
to study the dependence of the error $\delta E$ on the width $W$.  The
error $\delta E_m$ increases exponentially with growing width $W$
(figure \ref{fig:error}).

\subsection{The Phase Transition and the Critical Exponent.}

\begin{figure}
\begin{center}
  \epsfig{figure=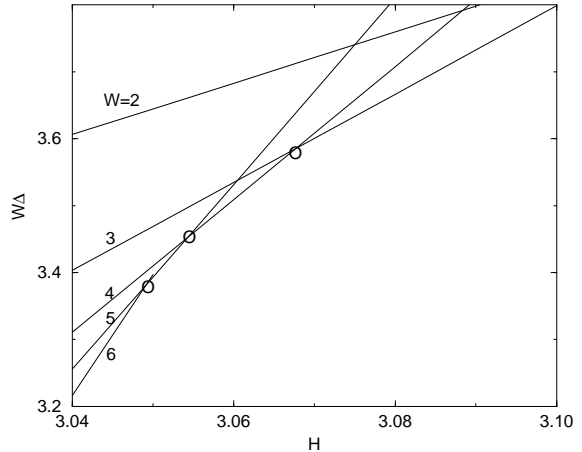,width=\linewidth}
\caption{The scaled gap $W\Delta$ as function of the field $H$ for
  aspect ratio $L=2W$. $W=2,3,4,5,6$. The curves become steeper with
  increasing width $W$. The crossings for consecutive widths are
  encircled. The system is periodical in both directions.  $m=128$,
  for $W=2,3,4,5$ and $m=200$ for $W=6$) }
\label{fig:Wgap2W}
\end{center}
\end{figure}

\begin{figure}
\begin{center}
  \epsfig{figure=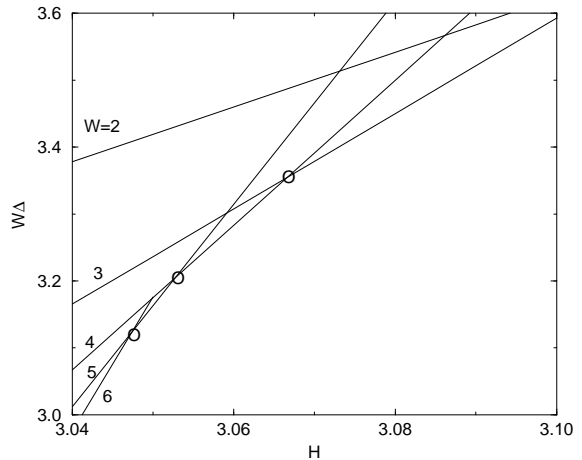,width=\linewidth}
\caption{Idem as figure \ref{fig:Wgap2W} with now the ratio $L=3W$.}
\label{fig:Wgap3W}
\end{center}
\end{figure}

The phase transition point $H_c$ is determined through equation
(\ref{eq:Hc}). We plot $W\Delta$ versus $H$ (figure \ref{fig:Wgap2W}
and \ref{fig:Wgap3W}). The curves would intersect precisely at $H_c$,
if it were not for corrections to scaling. These become quite large
when $W=2,3$. Afterwards we use formula (\ref{eq:nu}) to obtain $\nu$
at the intersection of the curves for consecutive widths $W$. The
results are listed in table \ref{tab:all}. For $W=6$ and $L=2W,3W$ we
are at the limit of our precision, when we take $m=128$ states. We
therefore set $m=200$ in this case.

\begin{figure}
\begin{center}
  \epsfig{figure=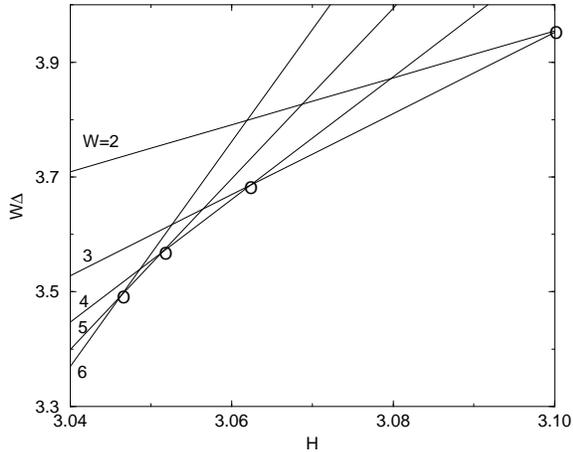,width=\linewidth}
\caption{The scaled gap $\Delta W$ for systems with open boundary 
  conditions in the length-drections and periodical in the
  width-direction. $L=5W$ and $m=64$.}
\label{fig:Wgap5W}
\end{center}
\end{figure}

Apart from these periodical systems, we have also considered systems
where the periodical connection between $A$ and $C$ is removed. The
removal of this boundary connection has two effects: First, the
accuracy of the calculated energies will increase substantially as the
size of the interacting boundary is halved. Second, the corrections to
scaling will increase. To make up for this second effect, we have to
resort to fairly large systems; $L=5W$. This is depicted in figure
\ref{fig:Wgap5W}.

From the values in table \ref{tab:all} we note that the corrections to
scaling for $\nu$ are still fairly large for these system sizes ($\sim
5\%$). We found that these corrections could not be compensated by
introducing an irrelevant scaling field in the relation
(\ref{eq:scal1}).

\begin{table}
  \begin{center}
\begin{tabular}{|c|c|c|c|c|c|c|}
  \hline & \multicolumn{2}{c|}{$L=2W$} & \multicolumn{2}{c|}{$L=3W$} &
  \multicolumn{2}{c|}{$L=5W$} \\ $W$ & $H_c$ & $\nu$ & $H_c$ & $\nu$ &
  $H_c$ & $\nu$ \\ \hline 2-3 & 3.113 & 0.74 & 3.110 & 0.73 & 3.101 &
  0.74\\ 3-4 & 3.068 & 0.69 & 3.067 & 0.68 & 3.062 & 0.69 \\ 4-5 &
  3.054 & 0.67 & 3.053 & 0.67 & 3.051 & 0.67 \\ 5-6 & 3.049 & 0.66 &
  3.047 & 0.65 & 3.046 & 0.66 \\ \hline
\end{tabular}
\caption{The phase transition point $H_c$ and the critical exponent $\nu$.
  We take $H_c$ to be the value where $ W \Delta(W^{-1})= (W+1)
  \Delta((W+1)^{-1})$.  $\nu$ is calculated through equation
  (\ref{eq:nu}). The first two aspect ratio's $L=2W,3W$ are with
  periodical boundary conditions, the last $L=5W$ is with open
  boundary conditions in the length-direction.}
\label{tab:all}
\end{center}
\end{table}


\section{Conclusion}

In this paper we have presented an adaption of the DMRG method to two
dimensional spins systems. We follow the route of adding complete
bands instead of single sites to the system. The latter was done by
Liang, Pang \cite{liang94} and White \cite{white96}. This modification
allows us to force a translational symmetry in the width-direction.
The advantage of implementing this symmetry is that a ground state
with specific translational properties can be targeted. Moreover, the
space in which the ground state is sought is reduced substantially.
This is especially useful in systems with Goldstone modes or similar
gapless excitation spectra where the lowest excitations belong to
different symmetry classes than the ground state.

The computational effort still remains similar to the approach of
adding single sites as the larger space of the band ($2^W$ instead of
$2$) is offset by three reductions: First, the ground state can be
written more compactly (a factor $W$ reduction). Second, we only need to
apply one operator ${\cal S}^x$ per boundary instead of $W$ operators.
Third, the sub-system (part A) grows with a full band instead of a
single site per step (factor $W$).

We have only considered systems of widths upto $W=6$. In models where
the total spin or the number of particles is conserved, we can go to
larger widths.

We observe that at criticality, the number of states $m$ needed for a
given accuracy $\delta E/E$ grows exponentially with the width $W$, in
full agreement with Liang and Pang \cite{liang94}. Moreover, we have
proven that far enough from the phase transition the method will
reproduce perturbation theory.

The procedure does not get stuck at local minima as was sometimes
experienced by White and Scalapino \cite{white97}.

The gap $\Delta$ we have calculated is a nice example of the use of
symmetry classes. The results for the critical properties, $H_c=3.046$
and $\nu=0.66$, are in reasonable agreement with the series expansions
of Pfeuty and Elliott \cite{pfeuty77} and with the cluster Monte Carlo
calculations of Bl\"ote \cite{bloete97}.

As yet, this method is not as accurate as the more traditional methods
like Monte Carlo simulations. The accuracy could be improved when a
larger width could be handled by including several hunderds of states.
At present this would require the use of a super-computer. Still it
has to be stressed that the DMRG can handle problems that are out of
reach of Monte Carlo simulations due to the ``sign-problem''.

{\bf Acknowledgement.} We thank H. W. J. Bl\"ote for guiding us
through the classical analogue and supplying the state-of-the-art
value of the critical field $H_c$. We are indebted to W. van Saarloos
for his continued interest in the problem and his suggestions for the
scaling analysis.

\bibliography{}  \ecols

\end{document}